\documentstyle[floats,prl,aps]{revtex}

\def\etal{{\it et. al.}}
\begin{document}
\draft
\preprint{CWRU-P1-96}
\title{Circles in the Sky: Finding Topology with the Microwave Background
Radiation}
\author{ Neil J. Cornish$^1$, David N. Spergel$^{2,3}$ and
Glenn D. Starkman$^1$}
\address{$^1$Department of Physics, Case Western Reserve University,
Cleveland, OH 44106-7079}
\address{$^2$ Princeton University Observatory, Princeton, NJ 08544}
\address{$^3$ Department of Astronomy, University of Maryland,
College Park, MD 20742}
\maketitle
\begin{abstract}

If the universe is finite and smaller than the distance
to the surface of last scatter, then the signature of the
topology of the universe is writ on the microwave background sky.
Previous efforts to search for this topology have focused on
one particular model: a toroidal flat universe.  We show how
both the high degree of spatial symmetry of this topology and
the integrability of its geodesics make it
unreliable as a paradigmatic example, and discuss why topology
on scales significantly smaller than the horizon are not ruled
out by previous analyses focussing on this special case.
We show that in these small universes the microwave background will
be identified at the intersections
of the surface of last scattering as seen by different
``copies'' of the observer.  Since the surface of last scattering
is a sphere, these intersections will be circles, regardless
of the background geometry or topology.
We therefore propose a statistic that is sensitive to all small finite
homogeneous topologies.
Here, small means that the
distance to the surface of last scatter is smaller than the ``periodicity
scale'' of the universe.

\vspace*{-0.3in}
\end{abstract}

\section{Introduction}

One of the goals of research in cosmology is to answer basic questions about
the universe: ``What is its structure?'' ``Is it infinite or finite?" ``Will
it last forever?'' and ``How will it end?''. In the context of general
relativity, these questions can be stated more formally as ``What is the
{\it geometry} and {\it topology} of the universe?''

If the universe is homogeneous and isotropic on large scales, then its {\it %
geometry} is determined entirely by $\Omega$, the ratio of the current
average energy density to the critical energy density. If $\Omega>1$, then
the geometry of the universe is positively curved, like the surface of a
sphere; the volume of the universe is finite; and, for most equations of
state, the universe will ultimately recollapse in a Big Crunch (though see
\cite{KamionToumb}). If $\Omega=1$, then the geometry is flat, like a sheet
of paper, and the universe will go on expanding forever, albeit at a
velocity that asymptotically approaches zero. Finally, if $\Omega<1$, then
the geometry is hyperbolic (negatively curved) like the surface of a saddle,
and the universe will go on expanding forever, at a velocity that does not
asymptotically approach zero.

{\it Geometry} constrains, but does not dictate, {\it topology}. If the
geometry of the universe is flat, then it can either be infinite or compact.
There are different compact universes associated with each crystal group:
for example, a three torus corresponds to a cubic symmetry. On the other
hand, if the geometry of the universe is positively curved ($\Omega > 1)$,
then the universe must be compact. Finally, if the universe is hyperbolic,
then again it can be either infinite or compact. There is a rich branch of
mathematics associated with the study of compact hyperbolic geometries\cite
{thurston}.

There are several physical and philosophical motivations for considering
compact universes. Einstein and Wheeler advocate finite universes on the
basis of Mach's principle\cite{Mach}. Others argue that an infinite universe
is unaesthetic and wasteful\cite{genetic} because anything that can happen
does happen, and an infinite number of times. Quantum cosmologists have
argued\cite{nucleation} that small volume universe also have small action
and are therefore more likely to be created. More intuitively, it is
difficult to produce a large universe, so it happens less often. Finally, a
common feature of many quantum theories of gravity is the compactification
of some spacelike dimensions. This suggests a dimensional democracy, in
which all dimensions (or at least all space-like dimensions) are compact,
and geometry distinguishes the large ones from the small ones.
Positively-curved dimensions remained at or collapsed to Planck scales in a
Planck time, while negatively-curved dimensions grew to macroscopic
proportions.

Most of the scant attention to non-trivial compact topologies in cosmology
has focused on the simplest non-trivial topology of the flat geometry: a
cube with opposite sides identified, {\it i.e.}~ a three-torus, $T^3$. While
the universe may be truly flat ($\Omega \equiv 1$, not just $\vert \Omega -
1\vert \ll 1$), flat manifolds are measure zero in the set of all possible
3-manifolds. Moreover, in flat universes there is no scale set by the
geometry, so the dimensions of the fundamental cell of the topology (the
radii of the torus) are arbitrary. It would be an unexpected and unnecessary
coincidence if one of those scales was exactly of order the horizon size
today. As we shall discuss below, flat topologies, and $T^3$ in particular,
have features that make them unsuitable as exemplars of the behaviour of
universes with general topology.

If $\Omega <1$, however, then there is a natural scale for the topology,
namely the curvature scale. Indeed, the compact topologies of $H^3$
(hyperbolic 3-d space) are classified by their volume in units of the
curvature scale. It has been shown\cite{minvol} that the volume of any
compact hyperbolic 3-manifolds is bounded below by $V_{{\rm min}%
}=0.00082\,R_{curv}^3$, and many explicit examples have been constructed
with small volumes. A collection of relatively simple topologies have been
constructed by identifying the faces of the four hyperbolic analogs of the
Platonic solids, the hexahedron, icosahedron, and two dodecahedra\cite{best}%
. These examples typically have volumes in the range $(4-8)\,R_{curv}^3$,
but other simple examples have volumes of as small as $0.94\,R_{curv}^3$\cite
{weeks}. It has also been shown\cite{thurston} that all three manifolds are
built of primitives which are homeomorphic (topologically equivalent) to one
of eight possible manifolds of constant geometry. Moreover, in a
well-defined mathematical sense, {\it most} three manifolds are homeomorphic
to manifolds of constant negative curvature, {\it i.e.}~ to topologies of $%
H^3$. 

Recently, we\cite{cornish} proposed a new model for a compact hyperbolic
inflationary universe. This model was motivated by observations that suggest
$\Omega <1$: determinations of the Hubble constant, $H_o\simeq 80{\rm km}/%
{\rm s}/{\rm Mpc}$, observations of large scale structure that imply that $%
\Omega _{nr}h\sim 0.25$\cite{peacock}, and stellar ages that appear to
exceed the age of the universe if $\Omega =1$\cite{hogan}. Here, $\Omega
_{nr}$ denotes the ratio of the energy density in non-relativistic particles
to the closure density, and $H_o=h100{\rm km}/{\rm s}/{\rm Mpc}$. Previous
attempts to construct hyperbolic inflationary models\cite{openinf,openinf2} 
assumed that the topology of the universe was that of the universal covering
group. They required there to be two epics of inflation, one of which
solved the initial homogeneity problem and one of which had
exactly the right amount of inflation to lead to the observed value of 
$\Omega$.
By assuming that the universe was hyperbolic and compact, we were able to solve
the large-scale isotropy and homogeneity problems as long as the volume of
the universe was not much larger than $R_{curv}^3$, where $R_{curv}$ is the
curvature scale.

\section{Pre-COBE Attempts to Detect the Finite Scale of the Universe}

%
In a multiply connected universe, many null geodesics start from the
position of an object and reach the present observer. Thus there will be
many images of each object, often called ghosts. Many authors have sought to
use this fact to limit the scale of the topology by searching for multiple
images of recognizable objects. Unfortunately this approach is complicated
by the fact that the different images will present the object at different
epochs in its evolution, at different distances, at different redshifts,
with different reddening factors, and from different perspectives. In all
but a flat geometry, the images will also be stretched or compressed
differently.

Nevertheless, the maximum distance $d$ up to which we would have been able
to recognize our own galaxy has been discussed by several authors (\cite
{SokShvar}, \cite{Fag48},\cite{Fag52}), who conclude that the topology scale
is greater than $15h^{-1}{\rm Mpc}$. Gott\cite{Gott} uses the Coma cluster
as a probe of topology: Coma is a particularly rich nearby Abell cluster
dominated by two distinctive giant galaxies instead of one -- NGC 4874 has
many companions while NGC 4889 has few. He deduces that the topology scale
is at least $60h^{-1}{\rm Mpc}$. Others (\cite
{SokShvar,DemLap,Lehoucq,Fetisova}) examine catalogs of clusters, and yet
others\cite{Gott} use superclusters like Serpens-Virgo as topological
probes. These studies imply that $200h^{-1}{\rm Mpc}$ is a lower limit on
the topology scale. Quasars provide yet another possible candidate for
studying ghost images and have been discussed by several authors\cite
{Fag52,Paal,DemLap,NarSesh} including some who claim to have observed
periodicity in the quasar redshift distribution \cite{Burbidge,Fang} at a
scale of $125h^{-1}{\rm Mpc}$. This they sought to ascribe to large scale
topology. Finally, Biesada has suggested\cite{Biesada} that the claimed
excess of antipodal pairs of gamma ray bursts could be due to cosmological
topology.

\section{Limits on Topology from the Cosmic Background Radiation}

Because of the difficulty in identifying ghost images of objects reliably,
it is probably preferable to look for topology using the CMBR. For one
thing, since all the CMBR photons were emitted at nearly the same epoch one
does not have the same problem of modeling the evolution of the fundamental
object.

Fluctuations in the microwave sky can be produced by quantum fluctuations
that have been stretched to macroscopic scales by inflation. The temperature
fluctuations are given in terms of fluctuations in the gauge-invariant
gravitational potential $\Phi $:\cite{Mukhanov}
\begin{equation}
{\frac{\Delta T}T}({\bf r})\simeq {\frac 13}\Delta \Phi (\eta _{rec},{\bf r}%
)\;.
\end{equation}
$\Phi $ here is given by
\begin{equation}
\Phi ^{\prime \prime }+2(a/\phi _0^{\prime })^{\prime }(\phi _0^{\prime
}/a)\Phi ^{\prime }-\Delta \Phi +2\phi _0^{\prime }({\cal H}/\phi _0^{\prime
})^{\prime }\Phi =0\;.
\end{equation}
In the above equation $\phi_{0}$ is the background inflaton field, $a$ is the
 scale factor and primes denote partial derivatives with respect to
 conformal time $\eta$. The Laplace-Beltrami operator $\Delta$ must be expanded
 in terms of the discrete momentum eigenmodes allowed by the topology.
As a result, the quantum field operator $\hat \Phi $ is expanded as a
sum over the eigenstates of the spatial manifold ${\cal M}$.
For a compact manifold ${\cal M}=\widetilde{{\cal M}}/\Gamma $ where the
simply connected universal covering space $\widetilde{{\cal M}}$ is one of $%
E^3$, $S^3$, or $H^3$, and $\Gamma $ is a discrete group of isometries of $%
\widetilde{{\cal M}}$. For many topologies, these eigenmodes can be though
of as the harmonics of the fundamental polyhedra whose faces are identified
to form the topology. The behaviour of the quantum field depends crucially
on the nature of the the periodic orbits of ${\cal M}$ as these determine
the stationary eigenmodes. In this regard, there is a fundamental difference
in the behaviour of quantum fields in the compact topologies of $S^3$ or $%
E^3 $ and the compact topologies of $H^3$. For $S^3$ and $E^3$ there are a
finite number of fundamental periodic orbits, and all periodic orbits are
stable. In particular, there is a periodic orbit of maximum length, which is
fixed by the scale of the topology. We shall discuss $H^3$ below.

Attempts to search for topology in the CMBR have focused almost exclusively
on flat topologies. For example, in a cubic $T^3$ toroidal universe the
eigenstates are given by $\cos ({\bf k}\cdot {\bf r})$ and $\sin ({\bf k}%
\cdot {\bf r})$ where ${\bf k}$ takes on the discrete values
\begin{equation}
{\bf k}=\left( {\frac{2\pi n_x}L},{\frac{2\pi n_y}L},{\frac{2\pi n_z}L}%
\right) \;,
\end{equation}
where $L$ is the topology scale. There is thus a minimum wave number, $%
k_{min}=2\pi /L$, and a maximum wavelength $\lambda _{max}=L$. Importantly,
the existence of a maximum wavelength means that there can be no
fluctuations on scales larger than the topology scale $L$. Scott {\it et al.
}\cite{SSS} looks for mode cutoff in the power spectrum of the two-point
correlation function in the two-year COBE data, for cubic $T^3$ and an $n=1$
inflationary model. They show that the minimum comoving scale of such a
topology is $2400h^{-1}$Mpc. Using a statistic which is sensitive to planes
of symmetry in the temperature fluctuation map,
de Oliveira-Costa \etal\cite{deOliveira} show that, for a rectangular $T^3$,
the limit on the smallest cell dimension from the 4-year COBE data is
$3000h^{-1}$Mpc at 95\% confidence level.

The null detections prompt claims that small universes are no longer an
interesting cosmological model\cite{SSS}. This claim is based on the
assumption that the crucial mode cut-off mechanism seen in flat and
spherical topologies is generic to all compact topologies. It isn't. In fact
this bound severly constrains only $T^3$. Even in the other compact
topologies of flat geometry, the longest periodic orbit is twice, three times
or even six times the length of one of the sides of the fundamental cell, so
that a
lower limit on the maximum wavelength, $\lambda _{max}>3000h^{-1}$Mpc, is a
limit of $L_{cell}>1500h^{-1}$Mpc, $L_{cell}>1000h^{-1}$Mpc or even
 $L_{cell}>500h^{-1}$Mpc, which is significantly less restrictive.

In $H^3$, the situation is completely different from $E^3$ or $S^3$ -- there
are an infinite number of periodic orbits, all of which are unstable. {\it %
Not only is there no cut-off in the allowed wavelengths, the number of modes
grows exponentially with wavelength.} This exponential growth in the number
of eigenmodes of larger period is a consequence of the positive
Kolmogorov-Sinai entropy of the system\cite{h2}.
Moreover, since all eigenmodes are
unstable, a universe that begins its life with only one or two fundamental
modes excited will quickly evolve into a quantum state where all eigenmodes
are approximately equally occupied, with a completely random distribution.
The randomness is a reflection of the Bernoullian nature of the classical
dynamics. The usefulness of this ergodic mixing property of compact $H^3$
topologies as a pre-mixer for inflation was extolled in our earlier
paper\cite{cornish}.
Here, we point out that the absence of a maximum wavelength
cut-off invalidates any attempt to exclude small universes a priori on the
basis of a loss in long wavelength power due to topology. Instead, we
propose a different mechanism to search for topology using the microwave sky.

\section{Generic Features of Topology}

Whether the geometry be flat or hyperbolic, there are certain characteristic
observational signatures of topology. The surface of last scattering is the
surface of a sphere of radius $R_{ls}\simeq 2cH_0^{-1}\simeq 6000h^{-1}$Mpc
from which the CMBR photons were
emitted. In most cosmological models, the microwave fluctuations on the
large angular scales probed by COBE are due to variations in the
gravitational potential at the surface of last scatter. Thus, DMR can be
thought of as mapping the gravitational potential along the inner surface of
a two sphere whose radius is $R_{ls}.$ There are potentially other sources
along the line of sight, but these are unimportant, except for a possible
contribution to the quadrupole from the decay of potential fluctuations, on
angular scales larger than 1/2 degree \cite{jungman}. If the physical
dimension of the universe is less than $R_{ls}$, then the sphere crosses
back on itself and self-intersects. The loci of self-intersections are
circles.

This is easist to visualize in $T^3$. If the sphere just fits inside the
box, then three pairs of points on the sphere: $\{(0,0,d_{ls}),(0,0,-d_{ls})%
\}$, $\{(0,d_{ls},0),(0,-d_{ls},0)\}$, and $\{(d_{ls},0,0),(-d_{ls},0,0)\},$
will have identical temperature fluctuations. If we orient our spherical
coordinate system so that it matches the orientation of the periodic box,
then this degenerate case predicts that $T(l=0^{\circ },b=0^{\circ
})=T(l=180^{\circ },b=0^{\circ })$, $T(l=90^{\circ },b=270^{\circ
})=T(l=0,b=0^{\circ })$, and $T(b=90^{\circ })=T(b=-90^{\circ }).$ If the
sphere is somewhat larger than the box, then there will be three pairs of
circles, each centered around the center of each face of the cube, that
share common temperatures. For example, one of the pairs of circles will
inscribe the North and South Poles of the coordinate system: $%
T(l,b=b_0)=T(l,b=-b_0)$. Thus fluctuations in the CMBR would be correlated
on circles of the same radii centered on different points on the sky. As we
shall discuss below, the existence of these correlated circles allows us to
search for the existence of topology in general, independent of the
particular topology in question. It is important to emphasize here that {\it %
the signature is not constant temperature along each circle, but identical
temperatures at identified points lying along pairs of circles}. For the $%
T^3 $ topology, the value of $b_0$ depends on the ratio of the periodicity
scale of the universe, $L$, to $d_{ls}$.

Previous attempts to detect topology in a finite universe have used
statistics that are only sensitive to $T^3$ topologies\cite{SSS,deOliveira}
{\it This signature, identified circles on the sky, is generic to all
topologies and geometries.} The intersection of two spheres (or a sphere
reflected through a symmetry plane with itself) is always a circle. Thus, at
the surface of last scatter, there will be pairs of identified circles of
equal circumference. Regardless of whether the background geometry is flat,
spherical or hyperbolic, the mapping from the surface of last scatter to the
night sky is a conformal map. Since conformal maps preserve angles, the
identified circles at the surface of last scatter would appear as identified
circles on the night sky. The relative angle between the identified circles
will depend on the geometry and topology of the universe, as will the number
of pairs of identified circles. If we are able to detect these circles, then
their position, number and size can be used to determine the geometry and
topology of the universe.

A second generic feature of topology is that it makes space globally
anisotropic. This can be understood quite simply in the case of a
three-torus in which looking along one of the axes brings you back around in
a closed loop, but looking off-axis makes you wind round and round the space
like the red strip around a barber pole. Thus, in a topology, there are
preferred directions. What is more surprising is that all but $T^3$ also
make space globally inhomogeneous. In most topologies, the identifications
of faces are made with twists (much like how a Mobius strip is constructed
from a length of ribbon). Thus translating by the topology scale causes a
rotation. Since the topologies violate isotropy, the mixing of translations
and rotations causes a violation of homogeneity.

Unlike other inhomogeneous cosmologies, such as Tolman-Bondi universes which
are locally inhomogeneous, these violations of homogeneity and isotropy are
not excluded. After all, we already know that the universe is weakly
inhomogeneous and anisotropic on large scales -- there is observable
structure. Similarly, in the topologically interesting cosmologies,
homogeneity and isotropy are violated only by the correlations between the
structure that we observe -- such as the fluctuations in the CMBR. In
principle, the locations of the identified circles can also be used to
determine the orientation and location of the observer within the topology.

Since the basic signature of topology is identified circles on the night
sky, we have developed a statistical tool to detect these circles which we
are in the process of applying to the DMR\ 4 year maps. We begin by
selecting two points on the night sky, $\overrightarrow{x},$ and
$\overrightarrow{y}.$ We draw circles of radius $R$ around each point and
consider all possible relative phases, $\phi ,$ between the two circles. We
define a statistic,
\begin{eqnarray}
 S(\overrightarrow{x},\overrightarrow{y},R,\phi )\equiv &&\sum_{i=1}^n\left[
 C(\mid \overrightarrow{x_i}-\overrightarrow{y_i}\mid )+N^2(\overrightarrow{
 x_i)}+N^2(\overrightarrow{y_i})\right] \nonumber \\
 &&-\sum_i(T(\overrightarrow{x_i})-T(\overrightarrow{y_i}))^2
\end{eqnarray}
where $\overrightarrow{x_i}$ denotes the locus of $n$ pixels separated from
$\overrightarrow{x}$ by distance $R$ and $\overrightarrow{y_i}$ denotes the
locus of pixels separated from $\overrightarrow{y}$ by distance $R.$ $%
C(\theta )$ is the two point correlation function of the signal plus noise
at separation $\theta $ and $N^2($ $\overrightarrow{x_i})$ is the
expectation value of the noise at pixel $x_i.$ The expectation value of $S=0$
when we average it over all circles in the sky. However, if the universe is
periodic, and we have selected the appropriate pair of circles, then the
temperature at point $\overrightarrow{x_i}$ and point $\overrightarrow{y_i} $
will differ only by the pixel noise. Thus, along that circle, $\langle
S\rangle =\;nC(0).$ If the universe is periodic, then we should be able to
detect a number of such circles.

The distribution of the $S$ statistic in this basic model can be obtained as
a function of the radius of the circle. We expect that for small $R,$ and
for the COBE sample, one will be unable to distinguish a detection from
statistical noise due to the large variance in $S$ for small pixel number.
We expect that the variance in $S$ will be proportional to $\sqrt{n}%
(C(0)+2N^2).$ Since the signal is proportional to $nC(0),$ the higher signal
to noise fourth year data is vital for the analysis. The distribution of
this statistic is likely non-Gaussian and Monte-Carlo simulations will be
needed to evaluate whether the COBE\ data has sufficient signal/noise to
discriminate between models.

The COBE 4-year map is hopefully not our ultimate map of the microwave sky.
NASA is now considering three different proposed MIDEX missions (MAP, FIRE,
PSI), any of which would represent a several hundred-fold improvement in
signal-to-noise over the DMR map. If selected, one of these missions would
fly in either 1999 or 2000. ESA is also considering building the
COBRA/SAMBAS mission, which if selected, will be launched in 2003. Its
capabilities will be similar to the proposed MIDEX missions.

The topological signatures should be easily detectable (if they are present)
in the higher resolution lower noise maps that will soon be available. If
one finds generic signals of topology, one may then be able to construct
templates of different topologies, and be able to identify the particular
topology in which we live, where we are within the topology and which way is
up.

When the topology scale is larger than twice the distance to the last
scattering surface, then we would see only the first view of the CMBR -- its
echoes would not yet have reached us. Thus, there would be no identified
circles on the sky. One might still hope that there would generically be a
feature in the mode spectrum at the topology scale. In the toroidal flat
topologies, and even in the hexagonal flat topologies the mode spectrum is
simply understood. A low momentum cutoff should affect the Fourier transform
of the two point correlation function of the temperature fluctuations. What
remains to be studied is to what extent the signal from these effects is
overwhelmed by cosmic variance. This will depend on the particular topology,
since what is relevant is not just the topology scale (which is the distance
one must travel so that points in space are mapped back into each other),
but the distance that one must travel before points in phase-space are
mapped back into each other. This can be several times the topology scale.
In the limit that the topology scale becomes much larger than the horizon
size, the topology will no longer be identifiable.

In the hyperbolic geometry the chaotic nature of trajectories makes the true
eigenmodes of the wave operator non-analytic and so new techniques must be
used to do the mode expansion. These techniques have been partially
developed by mathematicians studying quantum evolution on two-dimensional
hyperbolic surfaces.

\section{Conclusions.}

The possibility of non-trivial topology greatly widens and enriches the zoo
of possible cosmologies. The profound differences between $T^3$ and almost
all other possible global structures, mean that the stringent limits on $T^3$
from the existence of a quadrupole do not extend to other topologies. In
particular, in hyperbolic geometries, trajectories are chaotic and they have
no mode cutoff. We have suggested that for small universes the ideal
signal is to look for circles of identification in the microwave background
and have devised a statistic which does so.

If we do detect the signature of finite topology, its implications would be
profound and have great popular interest: we will learn that we live in a
finite universe.


\begin{references}

\bibitem{KamionToumb}  Kamionkowski, M. and Toumbas, N., preprint 1996,
submitted to PRL.

\bibitem{thurston}  W. P. Thurston and J. R. Weeks, Scientific American,
July '84, 108; W. P. Thurston, {\em The Geometry and Topology of 3-Manifolds}
(Princeton University Press, Princeton, 1978).

\bibitem{Mach}  A. Einstein {\em The Meaning of Relativity} (Princeton
University Press, Princeton, 1955), p. 108; J.A. Wheeler, {\em Einstein's
Vision} (Springer, Berlin, 1968).

\bibitem{genetic}  G.F. Ellis, Q.J.R.Astron. Soc. {bf 16}, 245 (1975).

\bibitem{nucleation}  D. Atkatz and H. Pagels, Phys. Rev. {\bf D25}, 2065
(1982); Ya. B. Zel'dovich and A. A. Starobinskii, Sov. Astron. Lett. {\bf 10}%
, 135 (1984); Y.P. Goncharov and A.A. Bytsenko, Astrophys. {\bf 27}, 422
(1989).

\bibitem{minvol}  R. Myerhoff, Comment. Math. Helv. {\bf 61}, 271 (1986).

\bibitem{best}  L. A. Best, Can. J. Math. {\bf 33}, 451 (1971).

\bibitem{weeks}  J. R. Weeks, Ph.D. thesis, Princeton University (1985). Ya.
G. Sinai, Dokl. Akad. Nauk SSSR {\bf 124}, 768 (1959).

\bibitem{cornish}  N.J. Cornish, D. Spergel and G.D. Starkman, ``Does Topology
Facilitate $\Omega <1$ Inflation?'' submitted to Phys. Rev. Lett.

\bibitem{peacock}  J. A. Peacock \& S. J. Dodds, Mon. Not. R. Astr. Soc.
{\bf 267}, 1020 (1994); M. Kamionkowski \& D.N. Spergel, Ap. J., 431, 1
(1994); A.R. Liddle, D.H. Lyth, D. Roberts, \& P.P.T. Viana, to appear in {%
Mon. Not. R. Astr. Soc.} (1995) (astro-ph/9506091).

\bibitem{hogan}  M. Bolte \& C.J. Hogan, Nature, 376, 399 (1995).

\bibitem{openinf}  M. Sasaki, T.Tanaka, K. Yamamoto, J. Yokoyama, Prog.
Theor. Phys., 90, 1019 (1993); K. Yamamoto, M. Sasaki \& T. Tanaka, ApJ,
455, 412 (1995); B. Ratra and P.J.E. Peebles, Phys. Rev. {\bf D52}, 1837
(1995), M. Bucher, A.S. Goldhaber and N. Turok, Phys.Rev. {\bf D52}, 3314
(1995).

\bibitem{openinf2}A. Linde, Phys.Lett.B351:99-104 (1995); 
A. Linde and  A. Mezhlumian,  Phys. Rev. D52, 6789 (1995).

\bibitem{SokShvar}  Sokoloff, D.D. and Shvartsman, V.F., Sov. Phys. JETP 39,
196 (1974).

\bibitem{Fag48}  Fagundes, H.V., in Proc. of the Fourth Marcel Grossmann
meeting on General Relativity., ed. R. Ruffini, Elsevier (Amsterdam), pp.
1559-1563 (1986).

\bibitem{Fag52}  Fagundes, H.V. and Wichoski, U.F., Astrophys. J. 322, L57
(1987).

\bibitem{Gott}  Gott, J.R. III, M.N.R.A.S. 193, 153 (1980).

\bibitem{DemLap}  Demianski, M. and Lapucha, M. MNRAS, 224, 527 (1987).

\bibitem{Lehoucq}  Lehoucq, R., Luminet, J.-P., and Lachieze-Rey, M. 1994
preprint.

\bibitem{Fetisova}  Fetisova, T., Kuznetsov, D., Lipovetskii, V.,
Starobinskii, A., and Olovin, R., Astron. Lett 19(3), 198 (1993).

\bibitem{Paal}  Paal, G. Acta Phys. Acad. Scient Hungaricae 30, 51 (1971).

\bibitem{NarSesh}  Narlikar, J.V. and Seshadri, T.R. Astrophys. J 288, 43
(1985).

\bibitem{Burbidge}  Burbidge, G.R., Astrophys. J. 154, 241 (1968).

\bibitem{Fang}  Fang, L.-Z., Chu, Y., Liu, Y., and Cao, Ch., Astron.
Astrophys 106, 287 (1982).

\bibitem{Biesada}  Biesada,M. preprint (1993).

\bibitem{Mukhanov}  Mukhanov, V.F., Feldman, H.A. and Brandenberger, R.
 Physics Rep. {\bf 215}  (1992).

\bibitem{SSS}  Stevens, D., Scott, D. and Silk, J. Phys. Rev. Lett. {\bf 71}%
, 20 (1993)

\bibitem{h2} N. L. Balazs and A. Voros, Phys. Rep. {\bf 143}, 109 (1986).

\bibitem{deOliveira}  de Oliveira-Costa, A., Smoot, G. and Starobinskii, A.,
preprint astro-ph 9501019 (1995), submitted to Ap. J.

\bibitem{jungman}  Jungman, G. Kamionkowski, M. Kosowsky, A. and Spergel,
D.N., to appear in {\it Phys. Rev. D}. 

\bibitem{bennett}  Bennett, C.L., et al. 1996, COBE preprint 96-01,
submitted to Ap.J. Letters.

\bibitem{lineweaver}  Lineweaver, C.M., et al. 1994, ApJ, 436, 425L
\end{references}
\end{document}